\documentclass[pra, 12pt, showkeys, eqsecnum]{revtex4}
\usepackage{graphics}

\newcommand {\Tr} {{\mbox{Tr}}}

\begin{document}
\title{Why quantum dynamics is linear\footnote[2]{From talks based on reference \cite{me80}.}}
\author{Thomas F. Jordan}
\email[email: ]{tjordan@d.umn.edu}
\affiliation{Physics Department, University of Minnesota, Duluth, Minnesota 55812}

\begin{abstract}
Quantum dynamics is linear. How do we know? From theory or experiment? The history of this question is reviewed. Nonlinear generalizations of quantum mechanics have been proposed. They predict small but clear nonlinear effects, which very accurate experiments have not seen. Is there a reason in principle why nonlinearity is not found? Is it impossible? Does quantum dynamics have to be linear? Attempts to prove this have not been decisive, because either their assumptions are not compelling or their arguments are not conclusive. The question has been left unsettled. There is a simple answer, based on a simple assumption. It was found in two steps separated by 44 years.
They are steps back to simpler and more compelling assumptions. A proof of the assumptions of the Wigner-Bargmann proof has been known since 1962. It assumes that the maps of density matrices in time are linear. For this step, it is also assumed that density matrices are mapped one-to-one onto density matrices. An alternative is to assume that pure states are mapped one-to-one onto pure states and that entropy does not decrease. In a step taken in 2006, it is proved that the maps of density matrices in time are linear. It is assumed, as in the earlier step, that at each time the physical quantities and states are described by the usual linear structures of quantum mechanics, so the question is only about how things change in time. Beyond that, the proof assumes only that the dynamics does not depend on anything outside the system, but must allow the system to be described as part of a larger system.

\end{abstract}

\pacs{03.65.Bz}
\keywords{nonlinear quantum mechanics, nonlinear Schr\"{o}dinger equation}

\maketitle

\newpage

\section{Introduction}\label{one}

Can we prove that quantum dynamics has to be linear? That the Schrodinger equation has to be linear? What assumptions are needed? Is there a reason in principle that quantum dynamics is linear? If not, quantum dynamics could be a linear approximation of a nonlinear theory; experiments might reveal small nonlinear effects. This possibility is too interesting to give up easily. We want to see if it can work. We will accept that it is impossible only if we see an undeniably convincing reason. 

Wigner and Bargmann proved\cite{WignerGroupTheory,Bargmann64} that quantum dynamics must be linear if it does not change absolute values of inner products of state vectors. Bialynicki-Birula and Mycielski\cite{BirulaMycielski} proposed a nonlinear Schrodinger equation that inspired precise experimental tests\cite{Shimony,ShullEtAl,GahlerEtAl}. Weinberg\cite{WeinbergNlqAnn,WeinbergNlqPrl} proposed a general nonlinear form of quantum mechanics, which led to more experimental tests\cite{Bollinger,Chupp,Walsworth,Majumder}. The seriousness of these proposals and their experimental tests shows that the Wigner-Bargmann proof of linearity was not decisive; its assumption was not accepted. A more recent idea that relativity requires quantum dynamics to be linear \cite{GisinHPA,GisinExample,GisinX3} has not been conclusive. The question has been left unsettled. This history is described in Sections II-IV.

There is a simple answer, based on a simple assumption. It was found in two steps separated by 44 years. They are steps back to simpler and more compelling assumptions. A proof of the assumptions of the Wigner-Bargmann proof has been known\cite{jordan62,Kadison,Hunziker,Simon} since 1962. This earlier step is reviewed in Section V. It assumes that the maps of density matrices in time are linear. For this step, it is also assumed that density matrices are mapped one-to-one onto density matrices. An alternative is to assume that pure states are mapped one-to-one onto pure states and that entropy does not decrease.

In a step taken in 2006, it is proved that the maps of density matrices in time are linear \cite{me80}. It is assumed, as in the earlier step, that at each time the physical quantities and states are described by the usual linear structures of quantum mechanics, so the question is only about how things change in time. Beyond that, the proof assumes only that the dynamics does not depend on anything outside the system, but must allow the system to be described as part of a larger system. The proof is described in Section VII. The assumption is discussed in Section VIII.

\newpage

\section{Proposals and tests for nonlinear quantum dynamics}\label{two}

\smallskip
\large
\noindent
\textbf{Is quantum mechanics a linear approximation \\ of a more fundamental nonlinear theory? \\ With small nonlinear effects not yet seen? \\ How might they be seen?}
\normalsize
\bigskip

One possibility, proposed by Bialynicki-Birula and Mycielski\cite{BirulaMycielski}, is a nonlinear Schrodinger equation
\begin{equation}
\label{nlse}
i\frac{\partial \psi }{\partial t} = H\psi  = [\frac{-1}{2m}\nabla ^2 + V - b \, ln(|\psi |^2 )]\psi . 
\end{equation}
The size of the nonlinearity is set by a small positive number $b$. The $ln$ function makes it simple to normalize wave functions and to describe a system composed of noninteracting subsystems. It gives the familiar relation between energy and frequency. And there are soliton solutions, wave packets that do not spread. Altogether, this is a very attractive candidate for a nonlinear quantum mechanics. A proposal \cite{Shimony} to test it with neutron interferometry was realized first with a two-slit experiment \cite{ShullEtAl} that gave
\begin{equation}
 b < 3.4\times 10^{-13} eV 
\end{equation}
and then with Fresnel diffraction \cite{GahlerEtAl} that lowered the limit to
\begin{equation}
 b < 3.3\times 10^{-15} eV. 
\end{equation}
This is wave mechanics pure and simple. It is about how a wave function behaves if it satisfies this nonlinear wave equation.

What makes the dynamics nonlinear is that the Hamiltonian depends on the wave function; the Hamiltonian depends on the state. For a wave function of position, this gives a nonlinear Schrodinger equation like (\ref{nlse}). For a spin described by Pauli matrices $\Sigma_1 , \Sigma_2 , \Sigma_3 $, it means the Hamiltonian can be a matrix like
\begin{equation}
\label{Hspin}
H = \epsilon \langle \Sigma_3 \rangle \Sigma_3 .
\end{equation}
The Hamiltonian is a matrix that depends on the state, because $\Sigma_3 $ is a matrix and the mean value $\langle \Sigma_3 \rangle $ depends on the state. The equations of motion for the mean values $\langle \Sigma_1 \rangle $, $\langle \Sigma_2 \rangle $, $\langle \Sigma_3 \rangle $ for this Hamiltonian are the same as for any Hamiltonian that is a number times a spin matrix,
\begin{eqnarray}
\label{Eqspin}
\frac{d}{dt}\langle \Sigma_1 \rangle  & = & -2\epsilon \langle \Sigma_3 \rangle \langle \Sigma_2 \rangle \nonumber \\
\frac{d}{dt}\langle \Sigma_2 \rangle  & = & 2\epsilon \langle \Sigma_3 \rangle \langle \Sigma_1 \rangle \nonumber \\
\frac{d}{dt}\langle \Sigma_3 \rangle  & = & 0
\end{eqnarray}
These equations of motion are nonlinear because the Hamiltonian contains a mean value. They describe precession with a frequency that depends on the state.

This is a representative case of Weinberg's nonlinear quantum mechanics \cite{WeinbergNlqAnn,WeinbergNlqPrl}, which brings in nonlinear dynamics for spin. Weinberg's framework is very broad. It lets ordinary linear quantum mechanics be changed in various ways. For example, the representation of physical quantities by linear operators can be changed. I looked at that possibility and decided it is not a fruitful way to go; for a spin it gives nothing new \cite{me59}. I concluded, as I think everyone has, that it is more reasonable to assume that at each time everything is the same as in ordinary linear quantum mechanics, so the question is just how things change in time. Making that assumption, and following a suggestion of Polchinski\cite{Polchinski} to make Weinberg's theory applicable to a system composed of noninteracting subsystems, I showed that Weinberg's theory is just the statement that the Hamiltonian depends on the state, that the Hamiltonian, like our example (\ref{Hspin}),  involves mean values as well as operators \cite{me64,me66}. Actually, this gives a bit more than Weinberg's theory; it includes some simple examples that Weinberg's theory can not describe \cite{me66}.

Weinberg's proposal inspired four precise spectroscopic tests \cite{Bollinger,Chupp,Walsworth,Majumder}. Two looked for the effect of a term like (\ref{Hspin}) in the Hamiltonian. The other two looked for the effect of a similar term for spin $3/2$. None found a nonlinear effect. They put limits down to
\begin{equation}
|\epsilon | < 1.6\times 10^{-20} eV \quad \quad \quad \quad \] \\
\vspace{-1.4cm}
\[ < 2\times 10^{-27} \mathrm{\, of \: binding \: energy \: per \: nucleon} 
\end{equation}
on the size of the nonlinear interaction.

Are these searches for something that doesn't exist? Is nonlinear quantum dynamics impossible? Does quantum dynamics have to be linear?

\newpage

\section{The Wigner-Bargmann proof of linear dynamics}\label{three}

\smallskip
\large
\noindent
\textbf{Can we prove that quantum dynamics has to be linear? \\
Yes, if we assume that} \\
\vspace{-1.1cm}
\begin{eqnarray*}
pure \: states & \leftrightarrow  & pure \: states \\
\psi  & \leftrightarrow  & \psi ' \; \; is \: one \: to \: one \\
and \; \; \; |\langle\psi |\phi \rangle |^2 & = & |\langle\psi '|\phi '\rangle |^2.
\end{eqnarray*}
\normalsize
\smallskip
 
That is what Wigner and Bargmann\cite{WignerGroupTheory,Bargmann64} showed.  They proved that the change from $\psi $ to $\psi '$ can be made with an operator that is either linear or antilinear. The product of two antilinear operators is linear, so if the change can be made in two similar steps, like the change over a time interval that can be split into halves, the operator must be linear. If we assume a group property, that the change over a time interval is the change over part of the interval followed by the change over the rest of the interval done the same way, and we assume that changes of probabilities in time are continuous, we get the linear Schrodinger equation \cite{meLinearOperators}.

These proofs were made long before the proposals and tests for nonlinear quantum mechanics. Why were theories developed and experiments done to look for something that had been proved impossible? I think the reason is that the assumption of
\begin{equation}
|\langle\psi |\phi \rangle |^2 = |\langle\psi '|\phi '\rangle |^2
\end{equation}
is not compelling. What does it say? States represented by $\psi $ and $\phi $ at one time go to states represented by $\psi '$ and $\phi '$ at a later time. Given the state represented by $\psi $ at the earlier time, the probability that if you make a measurement at the earlier time you will find the state represented by $\phi $ is the same as the probability that if you wait and make a measurement at the later time you will find the state represented by $\phi  '$. That's reasonable. But does it have to be true? Can you imagine how it might not be true? Bialynicki-Birula and Mycielski describe one way. Weinberg describes another. The Hamiltonian depends on the state is how I would say it. Why not?

\newpage

\section{Relativity?}\label{four}

In the discussion following Weinberg's proposal and its experimental tests there was a new idea.

\bigskip
\large
\noindent
\textbf{Can we prove that quantum dynamics has to be linear? \\
Does relativity require it? \\
Does nonlinearity imply signals faster than light?} \\
\normalsize 

\noindent I have seen three papers that argue it does. Noting\cite{me80} that two of them\cite{GisinHPA,GisinX3} do not apply to Weinberg's theory, I am interested in only one\cite{GisinExample}.

It works with the specific example of Weinberg's theory described by the Hamiltonian (\ref{Hspin}) and equations of motion (\ref{Eqspin}) that we have already seen. The solutions of these equations of motion are that $\langle \Sigma_3 \rangle $ is a constant and
\begin{eqnarray}
\langle \Sigma_1 \rangle (t) & = & \langle \Sigma_1 \rangle (0) 
cos(2\epsilon \langle \Sigma_3 \rangle t) - \langle \Sigma_2 \rangle (0) sin(2\epsilon \langle \Sigma_3 \rangle t) \nonumber \\
\langle \Sigma_2 \rangle (t) & = & \langle \Sigma_2 \rangle (0) cos(2\epsilon \langle \Sigma_3 \rangle t)+ \langle \Sigma_1 \rangle (0 )sin(2\epsilon \langle \Sigma_3 \rangle t).
\end{eqnarray}
Consider two different sets of initial conditions: In set (a)
\begin{equation}
\langle \Sigma_3 \rangle (0 ) = \pm 1.
\end{equation} 
Then
\begin{equation}
\langle \Sigma_1 \rangle (0) = 0 =  \langle \Sigma_2 \rangle (0),
\end{equation} 
because the sum of the squares can't be larger than $1$, so
\begin{equation}
\langle \Sigma_1 \rangle (t) = 0 = \langle \Sigma_2 \rangle (t).
\end{equation} 
The initial conditions (b) are that
\begin{equation}
\langle \Sigma_3 \rangle (0 ) = \pm \frac{1}{\sqrt{2}} = \langle \Sigma_1 \rangle (0).
\end{equation} 
Then $\langle \Sigma_2 \rangle (0)$ is $0$ and
\begin{equation}
\langle \Sigma_2 \rangle (t) = \pm \frac{1}{\sqrt{2}}sin(\pm \sqrt{2}\epsilon t) = \frac{1}{\sqrt{2}}sin(\sqrt{2}\epsilon t).
\end{equation} 
Even for $50-50$ mixtures of the $+$ and $-$ cases, the result is different for the initial conditions (a) and (b). That, it is argued, allows signals faster than light.

The argument\cite{GisinExample} uses two particles with spin $1/2$ in the state with total spin $0$. The particles move away from their common source in different directions to separated locations where their spins are measured, one by Alice and one by Bob. This is the much-used set-up that comes from Bohm's 1951 version of the 1935 argument of Einstein, Podolsky and Rosen \cite{Bohm1951,einstein35a}.
\bigskip

\Huge $\; \; \; \; \; \; \; \; \; \; \; \; \; \; \; \; \; \uparrow \hspace{-0.65cm} \circ \longleftarrow \; \; \; \; \; \; \; \; \; 
\longrightarrow \circ \hspace{-0.69cm} \downarrow $  
\bigskip
\normalsize

If Alice and Bob measure the components of their spins in the same direction, then when Alice gets $+$, Bob gets $-$, and when Alice gets $-$, Bob gets $+$. This is experimentally verifiable; you can watch it happen over and over.

Suppose now that when a spin comes to Bob, he runs it through the nonlinear dynamics described by the Hamiltonian (\ref{Hspin}) for a time $t$ and then measures the component of the spin in the $y$ direction. Suppose he does this for each of a long string of spins while Alice measures the $z$ component of each spin that comes to her. Alice gets $+$ half the time and $-$ half the time, on the average. The argument\cite{GisinExample} is that then $\langle \Sigma_3 \rangle $ is $1$ for half the spins that Bob receives and $\langle \Sigma_3 \rangle $ is $-1$ for the other half. The spins that come to Bob are described by the initial conditions (a). After he runs a spin through the nonlinear dynamics and is ready to measure the $y$ component, Bob has a spin in a state where $\langle \Sigma_2 \rangle $ is $0$. He gets $+$ half the time and $-$ half the time for the $y$ component.

Suppose that instead Alice measures the component of each spin in the direction $45^\circ $ between $z$ and $x$. She gets $+$ half the time and $-$ half the time. The argument\cite{GisinExample} is that then the spins that come to Bob are $+$ in the $45^\circ $ direction half the time and $-$ in the $45^\circ $ direction half the time. They are described by the initial conditions (b). After Bob runs a spin through the nonlinear dynamics and is ready to measure its $y$ component, he has a spin in a state where $\langle \Sigma_2 \rangle $ is $(1/\sqrt{2})sin(\sqrt{2}\epsilon t)$. He can get $+$ more than half the time or $-$ more than half the time, depending on his choice of $t$. The argument\cite{GisinExample} concludes that Bob can tell whether Alice is measuring spin components in the $z$ direction or in the $45^\circ $ direction. Alice can send a signal to Bob. Since there is no restriction on the distances and times, the signal can be faster than light.

It's not necessarily so. If there is not time for a signal at the speed of light between the measurements made by Alice and Bob, then moving observers will disagree about which measurement happens first. The $50-50$ probabilities for Alice's results are calculated from the state of total spin $0$ for the two spins, without considering what Bob does. The probabilities for Bob's results can be calculated the same way, without considering what Alice does. Then there is no signal \cite{me64,me71,CzachorDoebner}.

If there is time for a signal at the speed of light between the measurements made by Alice and Bob, so Alice can call and tell Bob what is coming, then the states of Bob's spins are prepared as the argument describes, and the probabilities for Bob's results are what the argument says. Then there is a signal, but it is not faster than light.

What can be said \cite{me71,CzachorDoebner}, and this is where the argument gets some traction, is that in the presence of the nonlinear dynamics and the absence of signals faster that light, there must be an abrupt change in Bob's results as gradual changes of distances or times change the separation between the measurements made by Alice and Bob from timelike to spacelike. Is this weird? Yes. Does it prove that this nonlinear dynamics is impossible? No. That it does not is shown particularly clearly in a delightful paper written as a fantasy about a fictional character who finds a way to detect the changes in Bob's results \cite{Kent}.

\section{Proof of the Wigner-Bargmann assumptions}\label{five}

There is a simple solution. It was found in two steps separated by 44 years. They are steps back to simpler assumptions. A proof of the assumptions of the Wigner-Bargmann proof has been known\cite{jordan62,Kadison,Hunziker,Simon} since 1962. A proof of the main assumption used in it was found recently\cite{me80}. We review the earlier step in this section. The recent step is described in Section VII and VIII.

\bigskip
\large
\noindent
\textbf{Can we prove that quantum dynamics has to be linear? \\
Yes, if we assume, or can prove, \\ 
that the change of density matrices is linear.} \\
\begin{center}
\vspace{-0.8cm}
If $\rho \rightarrow \rho ' $  is linear and one to one, \\
then pure states  $\leftrightarrow $  pure states \\
and $|\langle\psi |\phi \rangle |^2  =  |\langle\psi '|\phi '\rangle |^2 $.
\end{center}
\normalsize

The proof is simple. We assume the dynamics applies to all states. In an interval of time, it maps every density matrix $\rho $ to a density matrix $\rho '$. The assumption that the map is linear means that if $\rho_1$ and $\rho_2$ are mapped to $\rho_1'$ and $\rho _2'$, then
\begin{equation}
\label{eq:ld2}
\rho = p\rho_{1} + (1-p)\rho_{2},
\end{equation}
for a number $p$ between $0$ and $1$, is mapped to
\begin{equation}
\label{eq:se1}
\rho ' = p\rho_1' + (1-p)\rho_2'.
\end{equation}

We assume the map is one to one, so it has an inverse. The inverse map is linear; the proof that the inverse of a linear operator is linear\cite[Theorem 7.1]{meLinearOperators} applies with attention restricted to density matrices.

Pure states are mapped to pure states. A pure state can't go to a mixed state: if $\rho_1'$ and $\rho_2'$ in Eq.(\ref{eq:se1}) are distinct, so are $\rho _1$ and $\rho _2$ in Eq.(\ref{eq:ld2}); thus if $\rho '$ is for a mixed state, so is $\rho $. The inverse map also takes pure states to pure states, so the set of all pure states is mapped one-to-one onto itself.

For each vector $|\psi \rangle$ of length $1$, let $|\psi '\rangle$ be a vector of length $1$ such that $(|\psi \rangle \langle \psi |)'$ is $|\psi '\rangle \langle\psi '|$. For each density matrix $\rho $ there are orthonormal vectors $|\psi_j \rangle$ and positive numbers $p_j$ whose sum $\sum_j p_j$ is $1$ such that
\begin{equation}
\label{eq:se2}
\rho  = \sum_j p_j|\psi_j \rangle \langle \psi_j |.
\end{equation}
The linearity implies that
\begin{equation}
\label{eq:se3}
\rho ' = \sum_j p_j |\psi_j' \rangle \langle \psi_j' |.
\end{equation}
Since $\langle \psi_j' |\psi_j' \rangle$ is $1$,
\begin{equation}
\label{eq:se4}
\Tr[(\rho' )^2] = \sum_{jk} p_j p_k |\langle \psi_j' |\psi_k' \rangle |^2 \geq \sum_j (p_j)^2 = \Tr[\rho^2 ].
\end{equation}
The same result for the inverse implies that
\begin{equation}
\label{eq:se5}
\Tr[(\rho ')^2] = \Tr[\rho^2].
\end{equation}
Let $|\psi \rangle$ and $|\phi \rangle$ be vectors of length $1$ and let
\begin{equation}
\label{eq:se6}
\rho  = \frac{1}{2} |\psi \rangle \langle \psi | + \frac{1}{2} |\phi \rangle \langle \phi |.
\end{equation}
Then
\begin{equation}
\label{eq:se7}
Tr[\rho ^2] = \frac{1}{2} + \frac{1}{2} |\langle \psi |\phi \rangle |^2
\end{equation}
and
\begin{equation}
\label{eq:se8}
\rho ' = \frac{1}{2} |\psi '\rangle \langle \psi '| + \frac{1}{2} |\phi '\rangle \langle \phi '|,
\end{equation}
so Eq.(\ref{eq:se5} ) implies that
\begin{equation}
\label{eq:se9}
|\langle \psi '|\phi '\rangle |^2 = |\langle \psi |\phi \rangle |^2.
\end{equation}
Wigner and Bargmann will take it from there.

Linear maps of density matrices can also be used to describe processes where pure states are mapped to mixed states, different states are mapped to the same state, the map is not onto all states or, generally, the map has no inverse that applies to all states. An assumption is needed to separate these processes from dynamics described by the Schrodinger equation, which has an inverse for all states. For this proof we assumed that the map of density matrices is one to one.

An alternative is to assume that it is the set of all pure states that is mapped one-to-one onto itself, instead of the set of all density matrices. Then the inequality (\ref{eq:se4}) is obtained as before. If it is assumed that the entropy does not decrease\cite{PeresEntropy}, then
\begin{equation}
\label{eq:se10}
\Tr[(\rho ')^2] \leq  \Tr[\rho ^2].
\end{equation}
This implies Eqs.(\ref{eq:se5}) and (\ref{eq:se9}).

\section{What is linear}\label{six}

The result of quantum dynamics is the time dependence of mean values for Hermitian operators representing physical quantities. This includes the time dependence of probabilities, which are mean values for projection operators. The result is the same whether it is obtained from the Schrodinger picture or the Heisenberg picture. The mean value $\langle P \rangle$ for a Hermitian operator $P$ is $\Tr \left[ P \rho \right]$ where $\rho $ is the density matrix that represents the state. In the Schrodinger picture, the time derivative of $\langle P \rangle$ is
\begin{equation}
\label{eq:ld1}
\frac{d}{dt} \langle P \rangle = Tr[ P \frac{d\rho} {dt} ] .
\end{equation} The linearity of the map of density matrices means that for a density matrix $\rho $ that is a mixture described by Eq.(\ref{eq:ld2}), the time derivative is
\begin{equation}
\label{eq:ld3}
\frac{d\rho} {dt} = p\frac{d\rho_{1}} {dt} + (1-p)\frac{d\rho_{2}} {dt},
\end{equation}
so for a mean value
\begin{eqnarray}
\frac{d\langle P\rangle }{dt} & = & \frac{dTr[P\rho ]}{dt} = Tr[P\frac{d\rho }{dt}] \nonumber \\ & = & pTr[P\frac{d\rho_{1}} {dt}] + (1-p)Tr[P\frac{d\rho_{2}} {dt}].
\end{eqnarray}
This means that the time derivative of a mean value depends on the state in a linear way. It can not be a product of mean values as it is in the equations of motion (\ref{Eqspin}) for our example of Weinberg's theory. The time derivative of an operator can be a product of operators; the equations of motion in the Heisenberg picture can be nonlinear. But the time derivative of a mean value has to be a linear function of mean values \cite{me80}.

\section{Proof of linear maps for density matrices}\label{seven}

\bigskip
\large
\noindent
\textbf{Can we prove that the change of density matrices is linear? \\ 
Yes, if we assume that the system can coexist with another \\ without interaction.} \\
\smallskip 
\normalsize

\noindent
The proposition that quantum dynamics is described most generally by linear maps of density matrices was set out\cite{sudarshan61a,jordan61} in 1961. Now we have a proof of it. This is the recent step \cite{me80}. It is the conclusion of the chain of steps, the proof of the assumption of the proof of the assumption of the Wigner-Bargmann proof. The assumption we use to prove it is discussed in the next section. Here is the proof.

Suppose the system $S$ that we are considering is one of two separate systems $S$ and $R$. We consider the larger system composed of the two subsystems $S$ and $R$. We assume there can be something else, the other system $R$, and we assume that it makes no difference. We assume the dynamics of $S$ does not depend on anything outside $S$. Not on what happens in $R$, or on the state of $R$, or on any correlations of the state of $S$ with the state of $R$. We assume that $S$ can be described as part of the larger system of $S$ and $R$ combined, but we assume that the dynamics of $S$ does not depend on the situation of $S$ in the larger system.

Suppose the state of the larger system of $S$ and $R$ combined is represented by the density matrix
\begin{equation}
\label{eq:ld4}
\overline{\Pi} = p\rho_1|\alpha \rangle \langle \alpha | + (1-p)\rho_2|\beta \rangle \langle \beta |
\end{equation}
where $|\alpha \rangle$ and $|\beta \rangle$ are orthonormal vectors for $R$ that do not depend on the time and, as before, $\rho_1$ and $\rho_2$ are density matrices for $S$ and $p$ is a number between $0$ and $1$. The reduced density matrix $\Tr_R \overline{\Pi }$, which is the density matrix $\rho $ for $S$, is described by Eq.(\ref{eq:ld2} ). The probability $\langle P \rangle $ for a proposition represented by a projection operator $P$ for $S$ is the sum of joint probabilities
\begin{eqnarray}
\label{eq:ld5}
\langle P\rangle & = & \Tr_{SR}[P\overline{\Pi}] = \Tr_{SR}[P|\alpha \rangle \langle \alpha |\overline{\Pi}]  +  \Tr_{SR}[P|\beta \rangle \langle \beta |\overline{\Pi}] \nonumber \\ & = & \langle P|\alpha \rangle \langle \alpha |\rangle  +  \langle P|\beta \rangle \langle\beta |\rangle
 = p\Tr_S[P\rho_1 ]  +  (1-p)\Tr_S[P\rho_2]. 
\end{eqnarray}
Suppose a measurement is made on $R$ that distinguishes the states represented by $|\alpha \rangle$ and $|\beta \rangle$. The probability is $p$  that the result is $|\alpha \rangle$  and $1-p$ that the result is $|\beta \rangle$. If the result is $|\alpha \rangle$, the probability for the proposition represented by $P$ is $\Tr_S[P\rho_1 ]$, and if the result is $|\beta \rangle$, the probability for $P$ is $\Tr_S[P\rho_2 ]$. This can be verified experimentally by repeating the process of preparing the state represented by $\overline{\Pi}$, measuring to distinguish the states of $R$, and testing various propositions for $S$. The density matrices $\rho_1$ and $\rho_2$ describe physically distinct possibilities. The times of the events can be changed as long as the measurement on $R$ is early enough. The time dependence of $\rho_1$ and $\rho_2$ will account for changes in the results. The time derivative of the probability for the proposition represented by $P$ is $\Tr_S[Pd\rho_1 /dt]$ or $\Tr_S[Pd\rho_2 /dt]$ depending on the result of the measurement on $R$. Altogether, with the probabilities for both results being considered, the time derivative of the probability for $P$ is
\begin{equation}
\label{prob12}
p\, Tr_S[P\frac{d\rho_{1}} {dt}] + (1-p)Tr_S[P\frac{d\rho_{2}} {dt}] = Tr_S[P(p\frac{d\rho_{1}} {dt} + (1-p)\frac{d\rho_{2}} {dt})].
\end{equation}

The probability $\langle P\rangle$ for the proposition represented by $P$ is also
\begin{equation}
\label{eq:ld6}
\langle P\rangle =\Tr_{SR}[P\overline{\Pi}] = \Tr_S[P\Tr_R\overline{\Pi}] = \Tr_S[P\rho ].
\end{equation}
Its time derivative is $\Tr_S[Pd\rho /dt]$. This is always correct. It may be the only possibility at hand. For example, suppose the state of $S$ and $R$ combined is represented by the density matrix
\begin{equation}
\label{eq:ld7}
\Pi = p\rho |\alpha \rangle \langle \alpha | + (1-p)\rho |\beta \rangle \langle \beta | = \rho [p|\alpha \rangle \langle \alpha | + (1-p)|\beta \rangle \langle \beta |].
\end{equation}
The dynamics for $S$ must be the same for $\Pi$ as for $\overline{\Pi}$.

The time derivative of the probability for the proposition represented by $P$ is always $\Tr_S[Pd\rho /dt]$. There are situations where it must also be described by Eq.(\ref{prob12}) to fit observations of events in a larger system. The dynamics in $S$ can not depend on the situation of $S$ in a larger system. Therefore
\begin{equation} Tr_S[P\frac{d\rho} {dt}] = Tr_S[P(p\frac{d\rho_{1}} {dt} + (1-p)\frac{d\rho_{2}} {dt})]. 
\end{equation}
From this equality for various projection operators $P$, we conclude that
\begin{equation}
\label{eq:ld3}
\frac{d\rho} {dt} = p\frac{d\rho_{1}} {dt} + (1-p)\frac{d\rho_{2}} {dt}.
\end{equation}

\section{The assumption at the bottom}\label{eight}

What we need to assume is just that the system we are considering can coexist with another system without interaction, that the dynamics we are considering can be independent of something else in the universe, that the system we are considering can be described as part of a larger system without interaction with the rest of the larger system.

This is like the implicit assumption we make every day when we consider a part of the universe and assume the rest has no effect. We take that assumption for granted. Without it there would be no physics at all.

Actually that implicit assumption is stronger than the assumption we make here. We implicitly assume that \textit{everything} outside the considered system has no effect. Here we assume only that there is \textit{something} out there that has no effect. The limit to the validity of the implicit assumption comes from the part of the universe outside the considered system that has the largest effect. The limit on the validity of the assumption we make here comes from the part that has the smallest effect.

The assumption we make here is close in spirit to the idea of envariance\cite{Zurek05}, but the proof of Section VII does not use quantum entanglement. The proof does not even require quantum mechanics; it could be done in classical mechanics as well. The correlations involved are classical. The substance would be the same in classical mechanics; only the language would be different.

\bigskip
\large
\noindent
\textbf{Can we prove that quantum dynamics has to be linear? \\ 
Yes, if we assume that the system can coexist with another \\ without interaction.} \\
\smallskip 
\normalsize

\section*{ACKNOWLEDGMENT}

I am grateful to Wojciech Zurek for a suggestion that brought me back to this subject with my eyes open.

\bibliography{ncp}
\end{document}